\documentclass[conference]{IEEEtran}

\usepackage{cite}
\usepackage[numbers, square, comma, sort&compress]{natbib}
\usepackage[utf8]{inputenc}
\usepackage[usenames,dvipsnames,svgnames,table]{xcolor}

\ifCLASSINFOpdf
   \usepackage[pdftex]{graphicx}
   \DeclareGraphicsExtensions{.pdf,.jpeg,.png}
\else
\fi

\usepackage[cmex10]{amsmath}
\usepackage{array}
\usepackage{mdwmath}
\usepackage{mdwtab}
\usepackage{eqparbox}
\usepackage[tight,footnotesize]{subfigure}
\usepackage[unicode]{hyperref}
\usepackage{footnote}

\begin{document}

\bstctlcite{IEEEexample:BSTcontrol}

\newcommand\blfootnote[1]{%
  \begingroup
  \renewcommand\thefootnote{}\footnote{#1}%
  \addtocounter{footnote}{-1}%
  \endgroup
}

\title{On the Way to Future's High Energy Particle Physics Transport Code}

\author{
  \IEEEauthorblockN{Gábor Bíró, Gergely Gábor Barnaföldi, Endre Futó,\\
For the GeantV collaboration$^*$}
\IEEEauthorblockA{Institute for Particle and Nuclear Physics\\
		  Wigner Research Centre for Physics of the Hungarian Academy of Sciences\\
		  H-1121 Budapest, Hungary}
}

\maketitle

\begin{abstract}
High Energy Physics (HEP) needs a huge amount of computing resources. 
In addition data acquisition, transfer, and analysis
require a well developed infrastructure too. In order to prove new
physics disciplines it is required to higher the luminosity of the
accelerator facilities, which produce more-and-more data in the
experimental detectors. 

Both testing new theories and detector R\&D are based on complex
simulations. Today have already reach that level, the Monte Carlo
detector simulation takes much more time than real data collection. This
is why speed up of the calculations and simulations became important in
the HEP community.

The Geant Vector Prototype (GeantV) project aims to optimize the
most-used particle transport code applying parallel computing and to
exploit the capabilities of the modern CPU and GPU architectures as well. With the
maximized concurrency at multiple levels the GeantV is intended to be
the successor of the Geant4 particle transport code that has been used
since two decades successfully. Here we present our latest result on the
GeantV test performances, comparing CPU/GPU based vectorized GeantV 
geometrical code to the Geant4 version.\blfootnote{$^*$See the Appendix for the list of the collaboration members.}
\end{abstract}


\IEEEpeerreviewmaketitle

\section{Introduction}

High Energy Physics (HEP) requires a significant amount of computational resources at several layers of the data taking processes.
Detector upgrades of the experiments generate further needs for the computational capacity of hardware, software, and middleware levels. 
Since the increase of the CPU-clock frequency saturated a few years ago, developers turned to push more effort for the software-development side in order to exploit all the capabilities exist in the parallel architectures and many-core computing.

Early performance analysis of the detector and theory simulations have
been already explored the bottleneck of the numerical calculations, which found to be
the particle transport -- the most time-consuming part of the simulations. Due to
this reason the investigation of the possibility to rebuild the most widely
used particle transport framework, Geant4 (GEometry ANd Tracking), in a vectorized/parallelized way was 
started~\cite{artic:geant5, artic:detsim, artic:vecgeom, artic:geantvgpu, artic:hephpc, artic:parallintel, pres:conc9, pres:conc8, pres:conc7,  pres:conc6, pres:conc5, pres:conc4, pres:conc3, pres:conc2, pres:conc1, pres:gpucomp}.

Recent version is the Geant4 simulation framework, which is
used by most of the experimental collaborations of the Large Hadron
Collider (LHC, CERN) and many others all around the world. Its main
purpose is to simulate the passage of particles through matter using
Monte Carlo methods. The development of the original simulation package
started in 1994, and since 2012 studies for the possible successor, the Geant Vector
Prototype (GeantV) project has started~\cite{artic:geant5, artic:detsim, artic:geantv}.
The Geant development was always a worldwide collaboration and it has many
application areas outside of particle and nuclear physics, such as: space
science, nuclear reactor development, or medical/radiological fields.

The basic idea of a parallel version of these simulations lays on the
vectorized geometry calculations. The already vectorized GeantV code
executes single instructions on multiple data (SIMD) in parallel. A
relevant speedup has already achieved on CPU architectures which
supports vector instructions.

Nowadays, it is a popular trend to transform numerical codes to able to
run on the fast, General-Purpose computing on Graphics Processing Units
(GPGPU)~\cite{artic:gpgpu}. This fashion fits nicely to the strategy of GeantV development, since
geometrical computations with GPUs are pretty effective, especially in
cases where the size of the input to be processed is big compared to the
CPUs. In the case of HEP geometric calculations for all particles are
independent, thus it is reasonable to prepare the particle transport
code for such cases and implement the support for GPUs or other
many-core architectures. 

The VecGeom library (Vectorized Geometry) is the geometry package which 
was started to develop in the context of the Geant Vector Prototype
project in 2013~\cite{artic:detsim, artic:vecgeom}. By the time it will be finished, this became a
standalone library that provides a GPU support via the Computer Unified
Device Architecture (CUDA) compiler as well. This 
CUDA backend already exists in VecGeom, which supports only NVIDIA
devices at the moment. On the other hand we investigated how could we
increase the portability and the performance on multiple architectures,
with the usage of the Open Computing Language (OpenCL) framework~\cite{artic:cernsum}. 

In this paper first we give a general overview of the Geant framework,
then we introduce the generalization of the particle transport simulation.
Finally we show the speedtest results of VecGeom on CPU and GPU
architectures and present the comparison between them.

\section{Project goals}

We introduce the general aims of the GeantV project focusing the parallelization and vectorization.

\subsection{New generation of particle transport simulations}

\begin{figure}[h!]
\centering
  \includegraphics[width=0.51\textwidth]{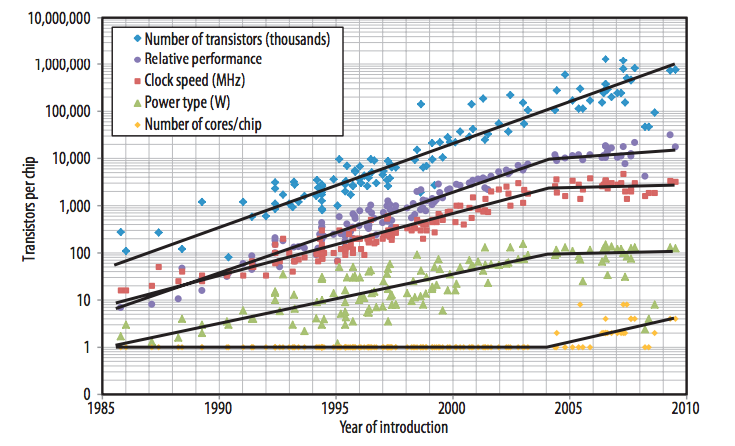}
  \caption{The famous Moore law says that the number of transistors integrated on one single chip grows exponentially with time~\cite{artic:moore}. However, it is still valid in hardware sense, but significant computational power has not been achieved so far due to the CPU-clock speed saturation on 3-4 GHz~\cite{fig:moore_fig}.}
  \label{fig:moore}
\end{figure}

The famous Moore law stated in the '60s says that the number of transistors integrated on one single chip grows 
exponentially with time~\cite{artic:moore}. The scaling law presented on Fig.~\ref{fig:moore}, is still valid 
in hardware sense. Although relevant speedup has not been achieved recently, because the clock frequency of 
the CPUs has reached the currently maximal value of 3-4 GHz since the middle of the last decade. Due to the 
increasing number of CPU cores per chip and other advanced architecture technologies the computing performance 
still able to be increased, but the exploitation of this increasing performance needs new, so far not widespread 
programming techniques. Since the popularity of video games increased recently, the graphical hardwares 
(GPUs) were greatly improved. Today these hardwares become suitable and used even for scientific 
calculations~\cite{artic:gpgpu, artic:vecgeom, artic:geantv, artic:parallpenel, artic:multicore, artic:katz}.

\begin{figure}[h!]
\centering
\includegraphics[width=0.49\textwidth]{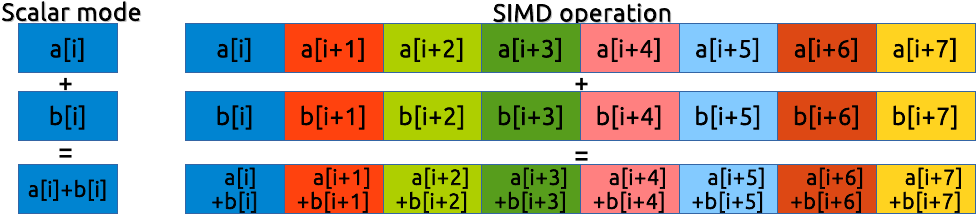}
\caption{{\sl Left panel}: A scalar code  performs one operation per CPU-tick, while a vectorized, SIMD-instruction based operation is capable to make the same operation on a data vector at the same time ({\sl right panel}).}
\label{fig:simdvsscalar}
\end{figure}

The power of the computing performance may not appear by increasing the clock frequency, but in the 
utilization of parallelization and vectorization. The parallel computing with CPUs has long 
history. The first SIMD instruction set appeared at late '90s, however in the last few years 
the new generations of CPUs have more-and-more advanced SIMD support (SSE4, AVX-512) and therefore 
they are becoming more effective with a vectorized software. The outline of the operating principle 
of the SIMD instructions is seen on Figure~\ref{fig:simdvsscalar}. 

\begin{figure}[h!]
\centering
\includegraphics[width=0.53\textwidth]{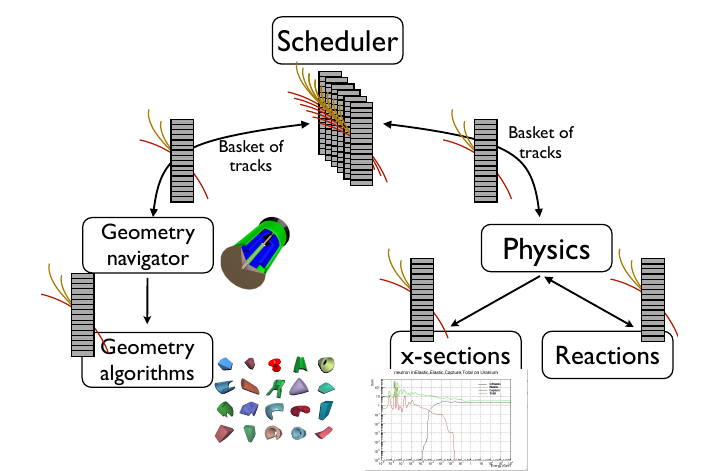}
\caption{The structure of the GeantV prototype: scheduler manages geometrical algorithms and physical processes~\cite{artic:geantv, artic:hephpc}.}
\label{fig:geantv_layout}
\end{figure}

The main goal of the Geant Vector Prototype Project (GeantV) is to exploit the benefits of the modern 
CPU, GPU and many-core architectures at the same time. The project started in 2012 with preliminary 
performance analysis which showed that in a typical detector simulation the geometry calculations itself using 
major of the computational resources, $40-50\%$ in time~\cite{artic:geant5}. On the other hand, geometry 
calculations are independent of each other, which gives an hand-on opportunity for parallelization.
The layout is seen on Figure~\ref{fig:geantv_layout}, where scheduler manages geometrical algorithms and physical processes~\cite{artic:vecgeom, artic:hephpc, artic:geantv}.

Development of the code aimed to get maximal performance beside maintainability and hardware independence.
On the top of this, the predecessor, Geant4 is developed and used since two decades and the successor 
has to be suitable for long term as well. Thus the software has to be compatible not just with the 
older hardwares but with future ones too. Following Fig.~\ref{fig:geantv_layout} the development of GeantV has three different levels: 

\begin{enumerate}
\item Scheduler: While in Geant4 the particle transportation happened serially, GeantV will manage multiple 
                 particles at the same time. Because the transportation is a local process and most of the 
                 simulation steps occur in a small part relative to the whole detector geometry, the main idea 
                 was to collect the particles which are in the same type of volume into vectors (baskets) and 
                 perform computations parallel using SIMD instructions. In this way both the necessary memory 
                 containing the different parts of the detector, and the necessary computational time can be 
                 strongly reduced. The scheduler manages the baskets of particles as an interface between the physics and geometry.

\item Physics: At present the physics of GeantV is tabulated, which means that instead of theoretical calculations 
               the physical processes are pre-calculated and ordered in multidimensional tables. Because in 
               this way all the possible final states are given in advance, the necessary computational time 
               is much less. On the other hand, because of the finite number of possible final states, the 
               precision will be less as well. For higher precision one should pre-calculate and load more final 
               states, requiring more memory. The physics of the final version of GeantV is still in the design phase.
 
\item Geometry: In a particle transport simulation a very significant part of the runtime (40-50\%) is devoted 
                to the geometrical calculations. Therefore it is important to take advantage most of the 
                computational capacities of the hardware in the implementation of the geometry code. The main 
                purpose of the VecGeom (Vectorized Geometry) package is to use more efficiently the different 
                hardware architectures via vectorization.

\end{enumerate}

\section{High Performance Computing in HEP}

In parallel to the development of detector- and data-acquisition technologies computing architectures went 
through a major a evolution during the last years too. In order to keep the scientific advancement, 
the necessary infrastructure needed to be improved following the construction of the huge particle 
accelerators and giant detector systems. This trend is still ongoing: the hardware manufacturers release 
the newer and better hardwares regularly -- sometimes motivated by the needs of the HEP community.

As we have mentioned earlier, in Geant4 in high-energy physics simulations the detector-particle interaction 
uses the main computational-resource part. Moreover any detector-simulation and -design require particle 
transport simulations especially, one need to calculate the detector efficiency or perform predictions 
by theoretical calculations. We note, besides HEP applications several other disciplines are also served 
such as medical applications or space sciences, where there is a need to simulate the transport of particles through matter.

In a transport simulation the most significant quantity regarding the computational time is the number of 
steps. While the particle passing the matter, it will interact with its environment many times, and the 
distance between two interactions called step length. For a given process, $i$ this length is determined 
by the mean free path, $\lambda_i$, which is calculated by the following formula:

\begin{equation}
\lambda_i = \frac{1}{n \sigma_i} \ \ ,
\label{eq:meanfreepath}
\end{equation}
where $\sigma$ is the total cross section of the given process and $n$ is the density of the medium. 
The total cross section is the sum of the cross sections for all processes:

\begin{equation}
\sigma_{total} = \sum_{i} \sigma_{i} \ \ .
\end{equation}

If the size of mean free path is large (e.g. in vacuum), the step length can be large as well, so one may needs only a few steps 
until one can eliminate the particle (when it leaves the range of interest). On the other hand if the mean free path is small, 
it means that interactions take place even in short distance. 
In this case one needs more steps with smaller $\lambda_i$ which require more computational time to simulate 
the same distance at the same precision.
The inverse of the mean free path equals to the inverse sum of the mean free path of all the possible interactions:

\begin{equation}
\frac{1}{\lambda_{total}}  = \sum_{i} \frac{1}{\lambda_{i}} \ \ .
\label{eq:meanfreepath2}
\end{equation}

The two main parts of a Geant4 particle transport simulation are physics and geometry.

\subsection{Physics}

For a given particle type at fixed energy it may happen different kinds of interactions. One can distinguish the 
following three types:

\begin{itemize}
    \item At rest: It occurs when the kinetic energy of the particle is zero. The selection criteria in the case of the 
    at rest processes is the lifetime of the particle (e.g. radioactive decay).
    \item Along step: Processes which happen during the transportation (e.g. ionization).
    \item Post step: The processes which take place after one step, after the transportation has been fully completed. This 
    depends on the interaction length (e.g. elastic scattering).
\end{itemize}

The process which is being implemented in the actual step is selected by the lowest mean free path. 
In the first step the code calculates all the possible mean free paths via Monte Carlo method for all the possible interactions at a given energy.
In the next step the shortest mean free path will be implemented. If the lowest mean free path is 
larger than the distance from the geometric border (which is the boundary layer of two volume with different materials), 
then the particle is being transported to the border. In the next step the mean free paths will be calculated with the 
material of the new volume.

These two methods (the calculation of the physical interaction length (\textit{GetPhysicalInteractionLength}) and the execution of 
the step (\textit{DoIt})) are implemented in every steps. During a simulation the tracking of a particle is ended when its energy 
becomes lower than a given threshold or it decays, or it leaves the investigated geometry. The execution of a step (transportation and 
the implementation of the processes) is calculated with Monte Carlo methods as well.
  
\subsection{Detector geometry}

Complex detector systems are built with basic three dimensional geometric units like rectangular solids and tubes. 
The fundamental geometric calculations, like coordinate transformations and distance calculations are implemented 
in the source code of these elementary bodies. Using boolean 
operations one can construct complex, composite elements as well. Finally with hierarchical connections one can build 
the whole detector geometry as an ordered system of the basic building blocks.
The geometric calculations, like e.g. distance calculations, do not depend on the particle type. 
This one of the most computational intensive tasks, since it has to be carried out in every step. 
Thus it is crucial to implement them in an efficient and fast way.
  
The time is determined by the number of steps. In Geant4 the simulation is serial, which means that 
it tracks all the particles individually, one after the other. This results that the running time of the simulation will 
be proportional to the number of the particles. 
Since the simulation of the large detector systems need a huge amount of resources, 
the next generation of particle accelerators and detectors 
require to develop the new generation of particle transport simulations. 
For example: the simulation of one single particle in the ALICE detector system needs a few $\sim$ms, while in the LHC there are 
particle collisions and particles to be detected in every $\sim$ns.

\subsection{VecGeom}
The VecGeom library is the vectorized geometry package of the GeantV which is devoted to speed up the 
code of the elementary three dimensional solids and the basic calculations: rotations, translations, 
logical operations, and distance calculations~\cite{artic:detsim, artic:vecgeom}. During a simulation there are three 
questions to be answered for all the particles:

\begin{itemize}
  \item Is the particle inside or outside the solid?
  
  \item What is the minimal distance between the particle and the border of the solid?
  
  \item What is the maximal step length for a particle with a given velocity and direction?
  
\end{itemize}

The purpose of VecGeom is to answer these questions as fast as possible with parallelized calculations. 
The main idea is to collect all the particles which are located in the same type of volume and perform 
the calculations for all of them simultaneously with SIMD operations. In addition, the same operations 
have to be executed with different architectures, while minimizing code duplication. 
Using of template classes the VecGeom library is able to run the same code on either CPUs or GPUs even 
with or without vector instructions.

For vectorization on CPU architecture one may have multiple choices. The VecGeom uses the Vc SIMD library to perform 
explicit vectorization~\cite{url:vc}. Thereby the code will be always vectorized regardless of the compiler. 
In order to use the same code on GPU architectures the VecGeom uses CUDA platform which works very 
nicely with the CPU vectorized C++ code too. 
A further option can be the OpenCL (and SYCL) version, which is under construction~\cite{artic:cernsum}.

\begin{figure}[t!]
\centering
\includegraphics[width=0.49\textwidth]{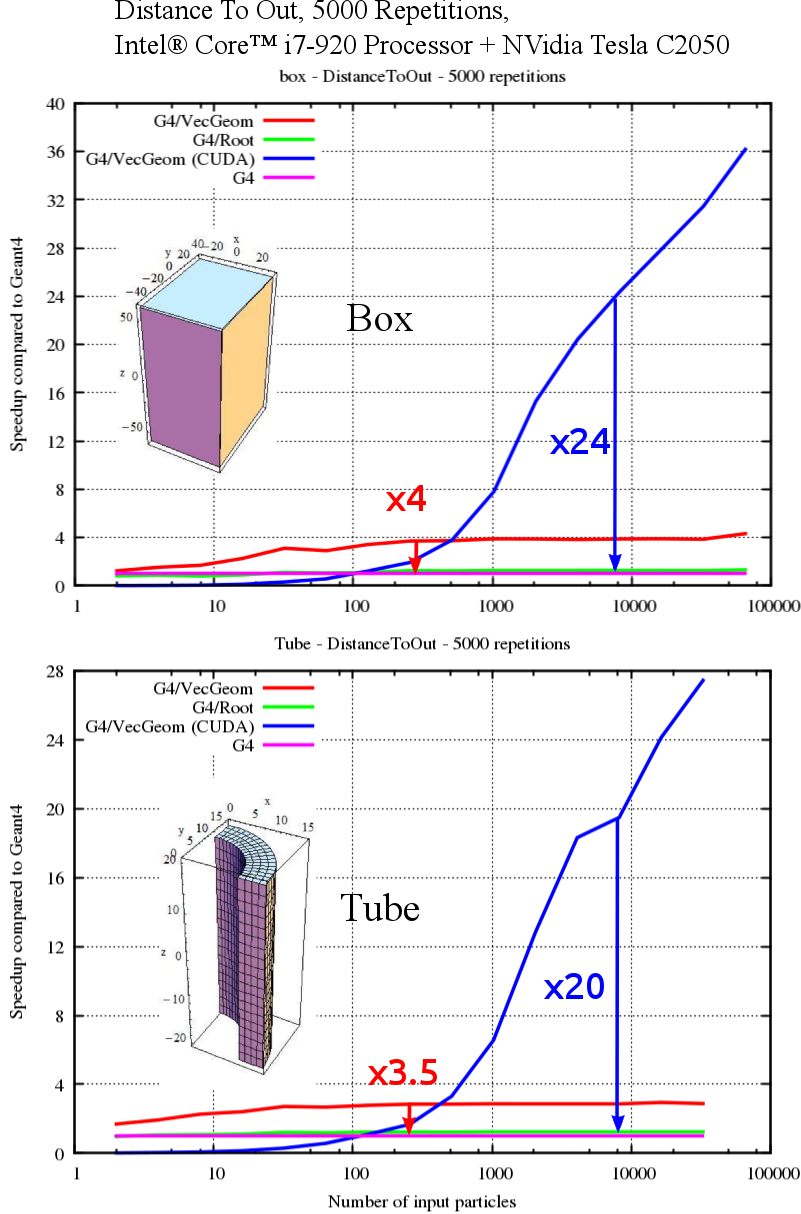}
\caption{Speedtest of the VecGeom \textit{Box} and \textit{Tube} bodies - the used hardwares are an Intel Core i7-920 CPU 
(SSE4.2 instruction set) and a NVIDIA Tesla C2050 computing processor~\cite{gpu:labor}.}
\label{fig:benchmark}
\end{figure}

In order to present the power of the VecGeom library, we performed a speed test of two bodies, a Box and a Tube, presented on Figure \ref{fig:benchmark}. 
During the test the speed of the DistanceToOut method were measured, which gives the distance of randomly placed particles with 
random speed escaping from the bodies. Fig. \ref{fig:benchmark} shows the relative speedup to the Geant4 calculations as a function of the 
number of particles and the vertical axis shows . In order to reduce the statistical 
fluctuations the calculations at a fixed particle number was repeated 5000 times.

The tendency of the tests is clearly visible: in case of CPU calculations a relevant speedup were measured even with low 
particle number, while in case of GPU to higher the particle number increase the speed of the calculations.

\section{Conclusion}
Since future HEP facilities require more-and-more speedup both by hardware and software way,  
the development of the next generation of GEometry ANd Tracking (Geant) code for many-core architectures was started. 
The aim is to build the vectorized GeantV code able to manage several times speedup calculations either on 
CPU and on GPU architectures. Here we presented the first promising results with the VecGeom library for simple 
body (Box and Tube) cases which were done in collaboration with the CERN PH-SFT group during a CERN Summer internship in 2014.

\section*{Acknowledgment}

This work was supported by Hungarian OTKA grants, NK106119, K104260, and
TET 12 CN-1-2012-0016. Author G.G. Barnaföldi also thanks the János
Bolyai Research Scholarship of the Hungarian Academy of Sciences. Author
G. Bíró also acknowledge the support by the Wigner Research Centre of
the H.A.S., the support of the Wigner GPU Laboratory and the support of the CERN-PH-SFT group.

\bibliographystyle{IEEEtran}
\bibliography{references}

\appendix[The GeantV Collaboration]
\label{app:geantv}
G. Amadio (UNESP), A. Ananya (CERN), J. Apostolakis (CERN), A. Arora (CERN), M. Bandieramonte (CERN), A. Bhattacharyya (BARC), C. Bianchini (UNESP), R. Brun (CERN), Ph. Canal (FNAL), F. Carminati (CERN), L. Duhem (intel), D.Elvira (FNAL), A. Gheata (CERN), M. Gheata (CERN), I. Goulas (CERN), F. Hariri (CERN), R. Iope (UNESP), S. Y. Jun (FNAL), H. Kumawat (BARC), G. Lima (FNAL), A. Mohanty (BARC), T. Nikitina (CERN), M. Novak (CERN), W. Pokorski (CERN), A. Ribon (CERN), R. Sehgal (BARC), O. Shadura (CERN), S. Vallecorsa (CERN), Y. Zhang (CERN)

\end{document}